\documentclass[conference, letterpaper]{IEEEtran}
\usepackage{cite}
\usepackage{amsmath,amssymb,amsfonts}
\usepackage{algorithmic}
\usepackage{graphicx}
\usepackage{textcomp}
\usepackage{xcolor}
\def\BibTeX{{\rm B\kern-.05em{\sc i\kern-.025em b}\kern-.08em
    T\kern-.1667em\lower.7ex\hbox{E}\kern-.125emX}}

\usepackage{tabularx} % for uniform column width

\usepackage{multirow}

\usepackage{geometry}
\usepackage[numbers]{natbib} %for arXiv
\usepackage{url}
\usepackage{xurl}

\begin{document}

\title {Enhancements for Developing a Comprehensive AI Fairness Assessment Standard\\
}

\author{\IEEEauthorblockN{Avinash Agarwal}
\IEEEauthorblockA{\textit{TEC, Ministry of Communications} \\
%\textit{Ministry of Communications}\\
New Delhi, India \\
https://orcid.org/0000-0003-4553-5861}
\and
\IEEEauthorblockN{Mayashankar Kumar}
\IEEEauthorblockA{\textit{TEC, Ministry of Communications} \\
%\textit{Ministry of Defence}\\
New Delhi, India}
\and
\IEEEauthorblockN{Manisha J. Nene}
\IEEEauthorblockA{\textit{DIAT, Ministry of Defence} \\
%\textit{Ministry of Defence}\\
Pune, India \\
https://orcid.org/0000-0003-0669-4464}
}
%Telecommunication Engineering Centre
%Defence Institute of Advanced Technology

\maketitle

\begin{abstract}
As AI systems increasingly influence critical sectors like telecommunications, finance, healthcare, and public services, ensuring fairness in decision-making is essential to prevent biased or unjust outcomes that disproportionately affect vulnerable entities or result in adverse impacts. This need is particularly pressing as the industry approaches the 6G era, where AI will drive complex functions like autonomous network management and hyper-personalized services. The TEC Standard for Fairness Assessment and Rating of AI Systems provides guidelines for evaluating fairness in AI, focusing primarily on tabular data and supervised learning models. However, as AI applications diversify, this standard requires enhancement to strengthen its impact and broaden its applicability. This paper proposes an expansion of the TEC Standard to include fairness assessments for images, unstructured text, and generative AI, including large language models, ensuring a more comprehensive approach that keeps pace with evolving AI technologies. By incorporating these dimensions, the enhanced framework will promote responsible and trustworthy AI deployment across various sectors.
\end{abstract}


\section{Introduction}
\label{introduction}
The widespread adoption of Artificial Intelligence (AI) and Machine Learning (ML) technologies has driven transformative advancements across critical sectors, including telecommunications, healthcare, finance, and public services. These intelligent systems now play a vital role in complex decision-making, pattern recognition, and process automation. This shift from traditional human-centered processes to AI-driven systems underscores the need to ensure that AI technologies operate in a fair and impartial manner. Unintended biases within these systems can lead to discriminatory outcomes, reinforcing prevailing societal inequities and undermining trust in AI applications \citep{ferrara2023fairness}.

As AI systems increasingly influence high-stakes decisions—such as eligibility for financial services, healthcare diagnoses, and law enforcement actions—demand for fairness and transparency in AI has never been more pressing. Biases in AI systems can arise from various sources, including biased training data, flawed algorithms, or unintended operational factors \citep{agarwal2023fairness}. These biases often produce disproportionately negative outcomes for marginalized or underrepresented groups, exacerbating social disparities and compromising the ethical credibility of AI. Therefore, addressing these biases and establishing reliable standards for fairness assessments have become priority areas for policymakers, industry leaders, and researchers \citep{agarwal2024seven}.

The telecommunications sector, in particular, is poised for further integration of AI, especially with the upcoming 6G era. As outlined in the ITU's IMT-2030 framework \citep{ITUIMT2030}, which includes \emph{AI and communications} as a usage scenario and emphasizes \emph{ubiquitous intelligence}, the future of telecom networks will increasingly rely on AI-driven functions like dynamic resource allocation, predictive maintenance, and hyper-personalized user services. However, as the role of AI in telecom networks expands, so does the risk of biased outcomes affecting network integrity, fairness, and trust. For example, in a 6G-enabled system, biased AI algorithms could inadvertently result in inequalities in network access, unfair allocation of network resources, skewed routing of network traffic, or other disparities \citep{umoga2024exploring} based on factors such as geography, device type, make-model of the network component, and environmental parameters. Furthermore, bias in services could create a scenario where marginalized communities might be further disadvantaged \citep{akter2021algorithmic}. Ensuring fairness in AI systems within the 6G context is essential not only to maintain network integrity but also to promote equitable access to advanced technologies and maintain user trust.

To address these challenges, the Telecommunication Engineering Centre (TEC) under India’s Department of Telecommunications has developed a specialized standard: the TEC Standard for Fairness Assessment and Rating of Artificial Intelligence Systems \citep{TECAISTD}. This standard provides a systematic framework to identify, assess, and measure fairness within AI systems, promoting more equitable and transparent AI applications. The TEC Standard employs a three-step assessment process: (a) conducting a bias risk assessment to identify potential fairness concerns, (b) establishing specific fairness metrics and threshold values to quantify fairness, and (c) performing comprehensive bias testing, including scenario-based evaluations, to ensure consistent performance across diverse user groups. This structured approach represents a significant advancement in enabling fairness in AI and promoting ethical AI practices. Unlike broader and high-level frameworks such as the National Institute of Standards and Technology’s (NIST) Risk Management Framework (RMF) \citep{NISTairmf}, which accommodates a variety of AI contexts, the TEC Standard offers a more detailed, implementation-level approach focusing exclusively on AI fairness. This specificity makes the TEC Standard a practical tool for developers and organizations to assess and ensure fairness in AI systems.

While the TEC Standard provides a solid foundation, it currently focuses on structured tabular data and supervised learning models. This limits its applicability as AI increasingly uses unstructured data and advanced models, such as Large Language Models (LLMs) and deep learning algorithms. Unstructured data, such as images and text, often lacks the structured labels of traditional datasets, making it more difficult to detect and mitigate biases that can arise in their interpretation and analysis. For example, biased image recognition systems can perpetuate harmful stereotypes or unfairly discriminate against certain groups \citep{fabbrizzi2022survey}. Similarly, unstructured text, such as social media content or news articles, can carry inherent biases in language, framing, or sentiment that influence AI decision-making \citep{bansal2022survey}. Large Language Models (LLMs), due to their vast training data, can inadvertently reinforce societal biases and generate harmful outputs, such as discriminatory language or biased content generation \citep{chu2024fairness}. These complexities highlight the need for targeted fairness assessments for these aspects not covered in the TEC Standard. 

Expanding the TEC Standard to encompass a broader range of data types and learning techniques would enable it to address fairness challenges in more complex AI applications, aligning with the rapidly evolving AI landscape. To address these needs, this paper proposes enhancements to the TEC Standard to accommodate the diverse data modalities and models prevalent in modern AI. Specifically, it recommends extending the standard to include fairness assessments for images, unstructured text, and generative AI models, such as LLMs. These extensions aim to ensure the TEC Standard remains relevant and effective across sectors with the advancements in AI technology.

This paper is organized as follows: Section 2 reviews the core elements of the TEC Standard, including its three-step process, multidimensional fairness view, and combined fairness metrics. Section 3 outlines our proposed enhancements for fairness assessments in images, unstructured text, and generative AI (LLMs). Section 4 discusses the implications of these enhancements. Section 5 concludes with insights on the potential of the TEC Standard to set a new benchmark in AI fairness.

\section{TEC AI Fairness Assessment Standard}
The TEC Standard for Fairness Assessment and Rating of Artificial Intelligence Systems \citep{TECAISTD} offers a structured framework for assessing and rating AI systems for fairness. The salient elements of this standard are as follows:
\subsection{Multi-Dimensional View of AI Fairness Assessment}
The TEC Standard adopts a multi-dimensional approach to assessing fairness in AI systems, aiming to provide a structured, comprehensive framework for identifying and mitigating biases as follows:
\subsubsection{Types of Bias} This dimension categorizes biases into three primary types: \emph{pre-existing biases} stemming from inequities reflected in the training data, \emph{technical biases} introduced during algorithm design or model training processes, and \emph{emergent biases} that arise when a model encounters new environments or operational contexts.
\subsubsection{Data Modalities} Although the TEC Standard centers on structured data, it recognizes the importance of various data modalities, such as text, images, audio, and video, in AI applications. Each modality introduces unique challenges in fairness assessment.
\subsubsection{Model Types} The TEC Standard recognizes that different machine learning models, supervised, semi-supervised, unsupervised, and reinforcement learning, require distinct fairness assessments. Additionally, model transparency (open, grey, or closed box) influences the selection of fairness evaluation methods.
\subsubsection{AI System Components} The TEC Standard promotes assessments across various AI system components, such as data, model, interfaces, pipeline, and infrastructure, as bias assessment algorithms and fairness metrics may vary by component, e.g., between data and model testing.
\subsubsection{Lifecycle Stages} Fairness assessments at multiple stages of the AI lifecycle, including data collection, model training, validation, and deployment, enable early bias detection and continuous monitoring. This approach ensures that fairness is integrated and newly introduced biases are identified as the system evolves.
\subsubsection{Risk Levels} Assessing the risk of bias in an AI system helps determine the required test data, data variation, and acceptable risk thresholds, considering the system's scope, nature, context, and purpose. The TEC Standard applies a risk-based approach to calibrate the fairness assessments.

\subsection{Three-Step Fairness Assessment Process}
The TEC Standard prescribes a systematic three-step process for fairness assessment, which includes:
\subsubsection{Bias Risk Assessment} This step involves identifying potential sources of bias in the AI system, including biases in training data, model architecture, or operational environments. The TEC Standard provides a detailed questionnaire to categorize risk levels - high, medium, or low, allowing for targeted risk management. This step focuses on uncovering biases that could lead to unfair outcomes, whether from societal preconceptions in the data or design choices in the model.
\subsubsection{Determination of Fairness Metrics and Thresholds} The next step establishes quantitative fairness metrics, such as Statistical Parity Difference (SPD) or Equal Opportunity Difference (EOD), to serve the AI system’s specific use case. Setting thresholds for these metrics ensures the system meets predetermined fairness standards, providing a clear, objective basis for fairness assessments across demographic groups.
\subsubsection{Bias Testing} The final step involves rigorous bias testing under various scenarios to detect potential fairness issues that could emerge during real-world use. Testing examines disparities in outcomes across protected attributes like race, gender, and age, ensuring the system aligns with fairness principles both pre- and post-deployment.

\subsection{Combined Fairness Metrics}
A distinctive feature of the TEC Standard is its inclusion of combined fairness metrics - Bias Index and Fairness Score, first introduced by \citep{agarwal2023fairness}. These metrics aggregate multiple underlying fairness metrics to deliver a comprehensive view of bias within an AI system. These metrics offer both attribute-specific and system-wide insights, enabling a more holistic fairness assessment that may not be immediately evident when examining individual fairness metrics separately.
\subsubsection{Bias Index} Bias Index quantifies bias across each protected attribute (e.g., race, gender) by integrating multiple fairness metrics and comparing outputs against ideal fairness values \citep{agarwal2023fairness}. This attribute-specific index provides nuanced insights into the extent of bias associated with individual characteristics, allowing for more targeted mitigation.
\subsubsection{Fairness Score} Fairness Score is an aggregate metric that synthesizes the Bias Index values of all protected attributes into a single composite score, offering a high-level overview of overall system fairness. A score near 1 indicates minimal bias and high fairness, while a lower score points to a higher bias level and less alignment with fairness standards \citep{agarwal2023fairness}. This metric enables stakeholders to evaluate AI fairness at a glance.

\section{Proposed Enhancements in the TEC Standard}
The TEC Standard’s current focus on fairness assessment in structured data and supervised models establishes a strong foundation; however, AI systems increasingly utilize diverse data modalities and complex architectures that present unique fairness challenges. The following enhancements extend the TEC Standard to accommodate fairness assessments for image data, unstructured text, and generative AI models.

\subsection{Fairness Assessment for Image Data}
With AI’s increasing reliance on computer vision (CV) in critical fields, fairness assessment for image data is essential to prevent discrimination. Image datasets used in sensitive applications, such as law enforcement or healthcare, can introduce biases that lead to harmful outcomes. Common sources of image bias include selection bias, framing bias, and label bias, each requiring targeted assessment to promote fairness. \emph{Selection Bias} occurs when datasets over- or underrepresent certain groups or contexts. This can skew model predictions, especially in sensitive applications like facial recognition, where overrepresented demographics may lead to more accurate predictions for those groups but higher error rates for others. Addressing selection bias requires statistical checks to ensure balanced demographic representation across the dataset, minimizing skewed outcomes. There may be however cases where selection bias may be essential. For instance, facial authentication that involves AI would seem to be an example where recognition should be biased toward a narrow group of people.\emph{Framing Bias} arises from image composition elements, such as camera angle, lighting, and cropping, which may subtly affect model interpretation. For instance, consistent lighting or angle biases might cause the model to associate certain attributes disproportionately with specific demographics. Fairness assessment methods should review compositional elements to verify that each demographic group is depicted in varied, representative ways, avoiding stereotypical portrayals. \emph{Label Bias} occurs when annotations introduce subjective or culturally loaded labels. Ambiguous labels (e.g., non-white) can obscure diversity within groups, leading to skewed predictions. Label bias can perpetuate stereotypes by associating certain appearances or attributes with specific demographics. Addressing label bias involves ensuring that labels are precise, objective, and consistent across the dataset, minimizing subjective categorizations \citep{fabbrizzi2022survey}.

The TEC Standard can include the following assessment techniques to address the biases in image data:
\subsubsection{Tabular Reduction Techniques} These techniques convert visual attributes into tabular form for statistical analysis, allowing fairness tests on demographic variables such as age and gender. This approach reveals representational imbalances across images, helping to identify hidden biases and ensure proportional representation of all demographic groups.
\subsubsection{Lower-dimensional Image Representations} These representations help visualize latent biases by examining clustering patterns in feature space. This method assesses if specific demographics are grouped or separated in ways that may indicate bias. By enabling visual analysis, it becomes easier to spot and address representational disparities.
\subsubsection{Cross-dataset Comparison} This technique provides an external validation check, assessing how well a model trained on one dataset performs across others. High performance on a single dataset but poor generalization may signal dataset-specific biases. “Name the Dataset” classifiers, which identify the source dataset for each image, further reveal such signatures, encouraging diversity across training datasets.
\subsubsection{Explainable AI (XAI) Methods} Techniques like saliency maps highlight which image regions most influence predictions, helping to identify if a model is based on relevant features or biased patterns. By pinpointing the focus areas of the AI model, XAI enables inspection into whether its predictions are justified and fair or whether it is influenced by irrelevant or biased features within images \citep{fabbrizzi2022survey}.

Integrating these methods into the TEC Standard can strengthen fairness assessments in image-based AI systems, promoting balanced and equitable representation in real-world applications. This comprehensive approach to fairness helps prevent biased predictions and reduces ethical risks in deploying AI models trained on image data.

\subsection{Fairness Assessment for Unstructured Text}
Fairness assessment in unstructured text data, such as online posts, articles, and dialogue, involves techniques to detect subtle forms of bias. Text data biases may manifest through language associations, implicit meanings, or representation imbalances, which are challenging to measure. Key methods for evaluating fairness in unstructured text are:
\subsubsection{Word Embedding Association Test (WEAT)} The WEAT is widely used to measure bias in word embeddings, which map words to vector spaces in NLP models. By evaluating the association between target words (e.g., professions such as engineer and nurse) and attribute words (e.g., gendered terms such as he and she), WEAT detects biases such as gender stereotypes. For instance, it can identify if professions traditionally associated with one gender (e.g., engineering) are more closely related to male terms. This method helps uncover societal biases embedded in word vectors, enabling refinement to reduce unfair associations \citep{gallegos2024bias}.
\subsubsection{Projection on Gender Directions and Distance Metrics} This approach quantifies bias by measuring a word's projection onto a gender direction (e.g., the vector difference between he and she). Words with a higher projection in this direction have a stronger gender bias. Distance-based methods also evaluate whether gender-neutral terms (e.g., nurse) are equidistant from male and female words in vector space. These techniques ensure that AI models do not inadvertently amplify gender biases in word representations, supporting more equitable language models \citep{bansal2022survey}.
\subsubsection{Sentence Encoder Association Test (SEAT)} The SEAT extends the WEAT to sentence-level analysis, where sentences rather than individual words are assessed for bias. It uses semantically neutral sentence templates, incorporating gendered or biased words to analyze their associations. SEAT enhances fairness assessments by considering the broader context in which words appear, ensuring that AI systems produce fair outputs in more complex text scenarios \citep{bansal2022survey, gallegos2024bias}.
\subsubsection{Gender Bias Evaluation Test Sets (GBETs)} GBETs are designed to evaluate bias in AI models by using gender-swapped versions of test data. This technique involves replacing gendered terms (e.g., he with she) and assessing the model performance on both versions of the dataset. A model free from gender bias would perform equally well on both gender-swapped datasets. GBETs are particularly useful for evaluating tasks like coreference resolution and sentiment analysis, where gender-based differences in predictions can indicate underlying biases.

Incorporating these fairness assessment practices in the TEC AI Fairness Standard will make it more comprehensive for text-based AI applications.

\subsection{Fairness Assessment in Large Language Models (LLMs)}
The increasing use of LLMs across diverse domains, from chatbots to healthcare, financial advisory, and legal aid, underscores their remarkable capabilities while highlighting the urgent need to ensure fairness in them. These models often inherit biases from real-world data, potentially exacerbating existing societal disparities. These biases can manifest in multiple ways, including discrimination based on race, gender, age, nationality, occupation, and religion. For instance, studies have revealed that LLMs like ChatGPT perpetuate gender stereotypes by associating leadership qualities more strongly with males than females, even when tasked with generating content like letters of recommendation \citep{gallegos2024bias}.

Identifying bias in LLMs is complex, as bias itself is subjective and influenced by various contextual and cultural factors. It often manifests in model outputs through representational biases, disparate performance, and reinforcement of harmful societal norms. These biases can lead to misclassifications or perpetuate negative stereotypes about social groups, further entrenching discrimination in AI systems. Techniques to assess bias in LLMs overlap with methods used for unstructured text but with additional layers specific to the architecture and nature of LLMs \citep{chu2024fairness}. These include:
\subsubsection{Bias Evaluation Using Language Model Outputs} LLMs can be tested by evaluating the generated text for imbalances or discriminatory patterns similar to bias detection in unstructured text. For example, tasks such as sentiment analysis can assess whether certain groups are portrayed more negatively or stereotypically.
\subsubsection{Bias Metrics in Embeddings} As LLMs rely on word and sentence embeddings, metrics, such as WEAT and SEAT, can be adapted to measure bias in LLMs. These tests can detect bias by assessing the association between gendered or racial terms and specific professions or traits. For instance, they can identify whether leadership traits are disproportionately linked to male-associated words.
\subsubsection{Probability-based and Generation-based Metrics} Probability-based metrics, such as masked token probabilities and pseudo-log-likelihood, assess biases by evaluating the likelihood of specific words or sentences in different contexts. These metrics can reveal underlying biases by comparing the model’s behavior when generating counterfactual pairs of sentences. Generation-based metrics assess bias by measuring output from biased prompts. Classifier-based metrics involve using auxiliary models to classify text generated by the LLM to detect differences in sentiment, bias, or toxicity, while distribution-based metrics evaluate the co-occurrence of biased or gendered terms in the model's outputs \citep{gallegos2024bias}.

Incorporating these techniques into the TEC Standard would provide a comprehensive framework for assessing biases in LLMs, ensuring that AI systems contribute to more fair and equitable outcomes across diverse applications.

\section{Discussion}
\label{discussion}
The proposed enhancements to the TEC Standard aim to address the growing complexity of AI systems across diverse domains. By expanding its scope to include fairness assessments for images, unstructured text, and generative AI (LLMs), the standard becomes more comprehensive and adaptable to the evolving landscape of AI applications.

Fairness assessments for images are essential as AI-driven computer vision technologies find increasing use in wide-ranging applications such as security, driverless cars, and medical diagnosis. Bias in image data can lead to significant mishappenings, such as wrong medical diagnoses or accidents of autonomous vehicles.

The inclusion of fairness assessments for unstructured text data is particularly relevant for social media, news content, customer service, recruitment, and more. AI systems often process vast amounts of unstructured text, and biases in this data can have far-reaching societal impacts. Ensuring fairness in these AI applications will help mitigate the risk of discriminatory outcomes, ensuring that they serve diverse populations without bias.

The focus on fairness assessment for generative AI (LLMs) responds to the increasing use of large language models in almost all sectors. These models can perpetuate harmful biases, particularly in high-stakes contexts. By introducing techniques like probability-based and generation-based metrics to assess bias in LLMs, the TEC Standard can help ensure that these systems do not reinforce stereotypes or contribute to discrimination in sensitive applications.

These enhancements will ensure that the TEC Standard remains relevant and robust in addressing fairness challenges as AI technologies become more pervasive across industries.

\section{Conclusion}
\label{conclusion}
The rapid adoption of AI across industries, including emerging technologies such as 6G and future networks, underscores the necessity for comprehensive fairness assessment frameworks. This paper reviewed the TEC AI Fairness Assessment Standard, highlighting its strengths and proposing essential enhancements to address the increasing complexity of AI systems. By expanding the standard to include fairness assessments for images, unstructured text, and generative AI (LLMs), a more robust framework is proposed that better reflects the evolving challenges of AI technologies.

These enhancements are crucial for ensuring that AI systems operate fairly, transparently, and accountably, particularly in high-stakes areas such as telecommunications, critical infrastructure, public services, healthcare, and transportation. These will help mitigate bias and discrimination in their outcomes and make them more reliable and trustworthy.

As AI continues to evolve, continuous updates and refinements to fairness standards will be necessary. The work presented in this paper represents a step forward in this ongoing effort, contributing to the broader goal of fostering trust and reliability in AI technologies.

\bibliographystyle{unsrt}
%\bibliography{references}

\begin{thebibliography}{10}

\bibitem{ferrara2023fairness}
{Ferrara, Emilio}.
\newblock {Fairness and bias in artificial intelligence: A brief survey of sources, impacts, and mitigation strategies}.
\newblock {\em Sci}, 6(1):3, 2023.

\bibitem{agarwal2023fairness}
{Agarwal, Avinash and Agarwal, Harsh and Agarwal, Nihaarika}.
\newblock {Fairness Score and process standardization: framework for fairness certification in artificial intelligence systems}.
\newblock {\em {AI and Ethics}}, 3(1):267--279, 2023.

\bibitem{agarwal2024seven}
{Agarwal, Avinash and Agarwal, Harsh}.
\newblock {A seven-layer model with checklists for standardising fairness assessment throughout the AI lifecycle}.
\newblock {\em {AI and Ethics}}, 4(2):299--314, 2024.

\bibitem{ITUIMT2030}
{Framework and overall objectives of the future development of IMT for 2030 and beyond}.
\newblock {\em Recommendation ITU-R M. 2160}, pages 1--19, 2023.

\bibitem{umoga2024exploring}
{Umoga, Uchenna Joseph and Sodiya, Enoch Oluwademilade and Ugwuanyi, Ejike David and Jacks, Boma Sonimitiem and Lottu, Oluwaseun Augustine and Daraojimba, Obinna Donald and Obaigbena, Alexander and others}.
\newblock {Exploring the potential of AI-driven optimization in enhancing network performance and efficiency}.
\newblock {\em {Magna Scientia Advanced Research and Reviews}}, 10(1):368--378, 2024.

\bibitem{akter2021algorithmic}
Shahriar Akter, Grace McCarthy, Shahriar Sajib, Katina Michael, Yogesh~K Dwivedi, John D’Ambra, and Kathy~Ning Shen.
\newblock Algorithmic bias in data-driven innovation in the age of ai, 2021.

\bibitem{TECAISTD}
{Fairness Assessment and Rating of Artificial Intelligence Systems}.
\newblock {\em TEC Std. 57050:2023}, pages 1--55, 2023.

\bibitem{NISTairmf}
{National Institute of Standards and Technology}.
\newblock {Artificial Intelligence Risk Management Framework (AI RMF 1.0)}.
\newblock \url{https://nvlpubs.nist.gov/nistpubs/ai/NIST.AI.100-1.pdf}, 2023.
\newblock Accessed: 09/11/2024.

\bibitem{fabbrizzi2022survey}
Simone Fabbrizzi, Symeon Papadopoulos, Eirini Ntoutsi, and Ioannis Kompatsiaris.
\newblock A survey on bias in visual datasets.
\newblock {\em Computer Vision and Image Understanding}, 223:103552, 2022.

\bibitem{bansal2022survey}
Rajas Bansal.
\newblock A survey on bias and fairness in natural language processing.
\newblock {\em arXiv preprint arXiv:2204.09591}, 2022.

\bibitem{chu2024fairness}
Zhibo Chu, Zichong Wang, and Wenbin Zhang.
\newblock Fairness in large language models: a taxonomic survey.
\newblock {\em ACM SIGKDD explorations newsletter}, 26(1):34--48, 2024.

\bibitem{gallegos2024bias}
{Gallegos, Isabel O and Rossi, Ryan A and Barrow, Joe and Tanjim, Md Mehrab and Kim, Sungchul and Dernoncourt, Franck and Yu, Tong and Zhang, Ruiyi and Ahmed, Nesreen K}.
\newblock {Bias and fairness in large language models: A survey}.
\newblock {\em {Computational Linguistics}}, pages 1--79, 2024.

\end{thebibliography}

\end{document}